\documentstyle[12pt,aasms]{article}

\def\etal{{\it et~al.\/ }}
\def\ref{\par\noindent\hang}

\def\etal{ et al.\ }

\def\Msun{\hbox{$\rm\thinspace M_{\odot}$}}

\def\ltsima{$\; \buildrel < \over \sim \;$}
\def\simlt{\lower.5ex\hbox{\ltsima}}
\def\gtsima{$\; \buildrel > \over \sim \;$}
\def\simgt{\lower.5ex\hbox{\gtsima}}

\begin{document}
\title{A GAUSS-HERMITE EXPANSION OF THE GALACTIC GLOBULAR
CLUSTER LUMINOSITY FUNCTION}

\author{ROBERTO G. ABRAHAM AND SIDNEY VAN DEN BERGH}
\affil {Dominion Astrophysical Observatory,
Herzberg Institute of Astrophysics\\
National Research Council of Canada\\
5071 W. Saanich Rd., Victoria B.C. V8X 4M6 Canada}

\begin{abstract} \noindent
We decompose the luminosity function of Galactic globular clusters into
a sum of the orthogonal Gauss-Hermite functions. This method quantifies
the asymmetric third-order ($h_3$) and symmetric fourth-order ($h_4$)
terms of the distribution while minimizing the effect of outliers in
the data. For 138 Galactic globulars we obtain $<M_V>=-7.41\pm0.11$,
$\sigma(M_V)=1.24$ mag, $h_3=0.02\pm 0.05$, and $h_4=0.06 \pm 0.05$,
{\em i.e.\/} the core of the distribution does not differ significantly
from a Gaussian.  For a low-metallicity subsample of 103 globular
clusters with $[Fe/H]<-0.8$, we find $<M_V>=-7.48\pm0.11$, $\sigma(M_V)=1.08$
mag, $h_3=0.05 \pm 0.05$, and $h_4=0.13 \pm 0.05$.
\end{abstract}


\section{INTRODUCTION}

\noindent The Galactic globular cluster luminosity function (GCLF) has
usually been represented by a Gaussian (log-normal) distribution, with
mean absolute magnitude $M_V \sim -7.3$ mag, and dispersion
$\sigma(M_V) \sim 1.1$ mag (Harris \& Racine 1979, Hanes \& Whittaker
1987, Racine \& Harris 1992, Secker \& Harris 1993). This
simple characterization provides a reasonably good description of the
GCLF expressed in terms of clusters per unit magnitude.
However van den Bergh (1985) has pointed out that the observed
luminosity function of globular clusters is slightly asymmetric, with a
long tail extending to faint absolute magnitudes. Recent studies have
suggested that the Galactic GCLF is non-Gaussian, and have attempted to
model it either as a t-distribution (Secker 1992; Racine \& Harris
1992), or to model the GCLF (expressed in terms of clusters per
unit luminosity) as a truncated power-law model derived by assuming a mass
function constructed from three power laws (McLaughlin 1993; Harris \&
Pudritz 1994).

In this paper we will adopt an alternative approach to characterizing
the Galactic GCLF. Since the luminosity function is quite close to a
Gaussian, we will decompose the observed GCLF into the sum of
orthogonal Gauss-Hermite polynomials. This technique was developed by
van der Marel \& Franx (1993), and independently by Gerhard (1993), to
describe the (continuous) line-of-sight velocity distributions of the
stars in galaxies. It was subsequently used by Zabludoff, Franx, \&
Geller (1993) to describe the (discrete) velocity distributions in a
sample of eight rich Abell clusters.  The expansion  into Gauss-Hermite
polynomials describes the overall distribution as the sum of a simple
Gaussian (the lowest order term in the expansion), along with symmetric
and asymmetric components whose relative sizes allow a precise
description of the departures from the simple Gaussian model.  An
additional benefit of this description is that the characteristic
parameters of the lowest-order Gaussian are insensitive to outliers in
the data. In the case of the Galactic GCLF, it is possible that
clusters in the faint tail of the distribution are dying objects such
as E3 (van den Bergh 1980), in which the high abundance of binaries
points to significant mass loss via evaporation.  By reducing the
sensitivity to these outlying points in our modeling, we hope to be
able to represent more accurately the intermediate-luminosity GCLF,
which may have been less affected by cluster evolution than the faint
tail of the complete luminosity function.

\section{METHOD AND RESULTS}

\noindent
Our decomposition of the luminosity function into a Gauss-Hermite
series is very similar to the well-known procedure for reconstructing a
function using Gram-Charlier series of type A (van der Marel \& Franx
1993; Kendall \& Stuart 1943). The major advantage of our
parametrization using Gauss-Hermite polynomials is that it is less
sensitive to outliers in the wings of the distribution than is the case
when higher order moments are used to characterize the data.  This is
often an important advantage in astrophysical applications.

Our technique is very similar to the procedure used by Zabludoff,
Franx, and Geller (1993).  Given a sample of $N$ absolute magnitudes,
$M_i$, $1<i<N$, the Gauss-Hermite moments $h_j$ of the luminosity
distribution are given by:

\begin{equation}
h_j  = {{2 \sqrt{\pi}} \over N \sigma(M_V)}\sum_{i=1,N}
        {{\exp(-x^2/2) \over \sqrt{2 \pi}} H_j(x_i)},
\end{equation}

\noindent where $H_j$ are the Hermetian polynomials given by:

\begin{eqnarray}
H_0(x) & = &1
\\
H_1(x) & = &\sqrt{2} x
\\
H_2(x) & = &{1 \over \sqrt{2}} (2 x^2 -1)
\\
H_3(x) & = &{1 \over \sqrt{6}} (2 \sqrt{2} x^3 -3 \sqrt{2} x)
\\
H_4(x) & = &{1 \over \sqrt{24}} (4 x^4 - 12 x^2 + 3).
\end{eqnarray}

\noindent The parameter $x_i$ is defined by:

\begin{equation}
x_i={{M_i-<M_V>}\over \sigma(M_V)}.
\end{equation}

\noindent For any choice of the parameters $<M_V>$ and $\sigma(M_V)$
the Gauss-Hermite series converges to the true function. In practise
one chooses $<M_V>$ and $\sigma(M_V)$ in such a way that the series
converges quickly, so that the lowest order term approximates the
series as well as possible.  As shown in van der Marel and Franx
(1993), choosing $<M_V>$ and $\sigma(M_V)$ so that the zeroth order
term is effectively the best-fitting Gaussian is equivalent to
solving for $<M_V>$ and $\sigma(M_V)$ in the system of equations:
\{$h_1=0; h_2=0$\}.  This choice of mean luminosity and dispersion for
the best-fitting Gaussian then fixes the third order (asymmetric) and
fourth order (symmetric) terms in the expansion, $h_3$ and $h_4$. This
expansion could, in principle, be extended to higher orders but we
have found that inclusion of terms up to the fourth order is sufficient
to adequately describe the observed luminosity distribution.

In order to apply this procedure to the analysis of the Galactic
globular cluster luminosity function, we used data from the recent
compilation of Djorgovski (1993) which assumes $M_V(RR)=+0.6$, and took
subsamples from this listing corresponding to (a) all clusters, (b)
low-metallicity clusters with $[Fe/H]<-0.8$, (c) clusters with
distances from the Galactic center, $R$, in the range $3<R({\rm
kpc})<30$, and (d) clusters with $[Fe/H]<-0.8$ and $3<R({\rm kpc})<30$.
The results obtained by decomposing the luminosity distributions of
these data using Gauss-Hermite polynomials are summarized in Table~1,
and illustrated in Figures~1--4. An estimate of the standard error of
$h_3$ and $h_4$ for $\sim 100$ objects has been given by Zabludoff,
Franx, \& Geller (1993) as $\sigma(h_3) \approx \sigma(h_4) \approx
0.06$. It should be emphasized that this estimate is a relatively crude
approximation and, like Zabludoff {\em et. al.} (1993), we rely on Monte
Carlo simulations to estimate the final uncertainties on these
parameters.  The confidence intervals from these simulations are shown
in Figure~5.  Note that the confidence intervals are {\em not} circular
(i.e estimates of $h_3$ and $h_4$ are correlated).  These correlations
appear to be somewhat stronger than predicted from the correlation
matrix given in van der Marel and Franx (1993), although the internal
correlations in this procedure are still much weaker than those
inherent in decomposition of a distribution into the sum of two
Gaussians.

The total sample of all 138 clusters deviates from a Gaussian at
barely the $1\sigma$ level, but The GCLF of the low-metallicity
($N=103$ objects) subsample deviates from the Gaussian with over
$2\sigma$ confidence. This suggests that the form of the GCLF might be
metallicity-dependent. However the evidence for this is rather weak,
since the overlapping $2\sigma$ confidence intervals for the complete
and low-metallicity subsamples seen in Figure~5 suggests that the two
luminosity functions differ at only the $\sim1\sigma$ level. This
agrees with the result obtained from a Kolmogorov-Smirnov test, which
suggests that these two distributions do not differ significantly.  In
order to test whether the observed values of $h_3$ and $h_4$ are
sensitive to the value $M_V(RR)=+0.6$ assumed by Djorgovski (1993)
in tabulating his globular cluster sample, we repeated the
calculations of $h_3$ and $h_4$ for cluster sample assuming
$M_V(RR)=0.30[Fe/H]+0.94$ as advocated by Sandage (1993), and
$M_V(RR)=0.15[Fe/H]+1.01$ as found by Carney \etal (1992). We find that
the assumption of a metallicity-dependent $M_V(RR)$ affects the
parameters $h_3$ and $h_4$ at the $\Delta h_3=\Delta h_4 =0.01$ level,
and is thus negligible in comparison with the confidence intervals
shown in Figure~5.

Since the analysis procedure described in this work is only suited to
the study of samples of $N \simeq $ 100 objects, probing finer
subdivisions within this sample is difficult. A sample of
low-metallicity clusters with $3<R({\rm kpc})<30$ ($N=71$ objects)
appears to deviate from a Gaussian at about $1.5 \sigma$ confidence.

\section{DISCUSSION}

\noindent The results presented in the previous section suggest that
deviations from the canonical log-normal GCLF might be most pronounced
amongst the low-metallicity subsample of the Galactic globular cluster
population. The dominant higher order component in our expansion of the
low-metallicity GCLF is the symmetric $h_4$ term. However, there is
also evidence for the presence of a small asymmetric contribution from
a non-zero $h_3$ term. The size of the $h_3$ and $h_4$ terms are small
enough that the overall low-metallicity luminosity function remains
quite well described by a Gaussian.  It is interesting to compare these
results with those obtained by Secker (1992), who investigated a
subsample of 100 globular clusters from the tabulation of Harris et.
al. (1991). This sample included globulars with Galactocentric
distances in the range $2<R({\rm kpc})<35$ that have reddenings
$E_{B-V} < 1.0$. These constraints on $R$ and $E_{B-V}$ were chosen in
order (1) to select a sample of globulars that would allow comparison
with the globular cluster population in M31, and (2) to be insensitive
to incompleteness caused by absorption.  Secker (1992) concludes that a
Gaussian fits the GCLF rather well (it cannot be excluded at the
2$\sigma$ level), but that a t-distribution gives a slightly better fit
to their data. This is partly due the extra degree of freedom in the
t-distribution model [the value of the likelihood function for the
best-fitting Gaussian and t-distribution models given in Secker (1992)
are very similar], but Secker's overall conclusion that the GCLF is
slightly non-Gaussian is quite consistent with our results. Secker's
preferred t-distribution has slightly wider wings than an equivalent
Gaussian, in good agreement with our detection of a small positive
$h_4$ term in the GCLF.  In our analysis we have made no attempt to
exclude globulars on the basis of visibility to an external observer.
Furthermore,  our sample includes a number of distant Palomar-type halo
clusters and bulge clusters that were excluded by Secker in order to
avoid skewing his model fits. The use of Gauss-Hermite polynomials in
modelling the GCLF allows the total sample (including outliers) to be
used without having to discard any data, which is a major advantage in
this context.  Since the Central Limit Theorem  (Kendall \& Stuart
1979) suggests that many observed processes in nature will have
underlying distributions that are Gaussians, we feel that there is
considerable appeal in a technique that allows one to accurately model
the GCLF as a Gaussian with small perturbations, rather than as a model
distribution chosen without a close connection to an underlying
physical or statistical process.

In an interesting paper, Harris and Pudritz (1994) have recently
attempted to model the observed luminosity function of globular
clusters (expressed in terms of clusters per unit luminosity) by using
a discontinuous multiple-component power-law model for the globular
cluster mass spectrum (under the assumption of constant mass-to-light
ratio).  This simple model is appealing because the shape of the
bright end of the globular cluster mass spectrum appears to be similar
in form in giant and supergiant galaxies. This shape can be
approximated by a two component power-law model with spectral index
$\alpha=1.6$ to $1.7$ between $10^{5.2}$ and $10^{6.5}\Msun$,
steepening to $\alpha=-3.2$ beyond $10^{6.5}\Msun$. Harris and Pudritz
(1994) and McLaughlin (1993) show that the mass spectrum of giant
molecular clouds can also be approximately described by a
two-component power-law model with spectral index $\alpha\sim 0.15$
between $10^4-10^{5.5} \Msun$, and $\alpha \sim 1.7$ at higher masses.
Harris and Pudritz argue that if the rate of globular cluster
destruction is negligible, and if the progenitors of globular clusters
are dense gaseous cores embedded in giant molecular clouds and HII
regions with mass spectra similar to those seen in giant molecular
clouds at the current epoch, then a ``universal" discontinuous
three-component power-law mass-spectrum can be defined between $10^4
\Msun$ and $10^{6.5} \Msun$ by simply simply joining the molecular
cloud/HII region mass spectrum with the globular cluster mass spectrum
seen in giant and supergiant galaxies. The GCLF expressed in terms of
clusters per unit magnitude resulting from this universal
mass-spectrum model is approximately Gaussian. While Harris and
Pudritz' model has the disadvantage of being discontinuous in the mass
spectrum, and has a relatively large number of parameters, the good
agreement between the near-Gaussian GCLF predicted by their model and
the observed near-Gaussian GCLF is intriguing. However if the weak
($\sim1\sigma$) evidence presented in the previous section for a
metallicity dependence in the GCLF is confirmed, this would be a
potentially serious problem for this model, since a
metallicity-dependent GCLF seems difficult to reconcile with Harris \&
Pudritz' assumption of a universal mass spectrum at all epochs.
Furthermore, is not clear that dense gaseous cores embedded in giant
dust-free metal-poor gas clouds in the turbulent halo of the
proto-Galaxy will have the same mass spectrum as that presently
exhibited by the dusty molecular clouds in the Galactic disk. It is
also not clear that the masses of globular clusters should be
proportional to those of their parent clouds.  Finally, a potentially
serious difficulty may lie in the assumption of negligible cluster
destruction. Calculations by Hut \& Djorgovski (1992) suggest that
$\sim 4\%$ of the total Galactic globular cluster population is
destroyed per Gyr as a result of relaxation-driven evaporation and
shocking by the combined effects of interactions with the Galactic
disk and bulge.  We note in passing that a Schechter function provides
an excellent fit to the Harris \& Pudritz universal mass-spectrum
model, except at the very low-mass end of the spectrum.  A primordial
Schechter-law mass-function, combined with systematic destruction of
low-mass globulars, might result in a present-epoch mass-spectrum
similar in form to the Harris and Pudritz three power-law model.

If cluster destruction is occuring, the small deviations from the
canonical Gaussian distribution measured by $h_3$ and $h_4$ can be
used to gain some insight into the shape of the primordial GCLF. In
simple models, wherein the cluster halo is formed through an
amalgamation of distinct (Gaussian) cluster populations in an early
merger, observed $h_3$ and $h_4$ values may restrict the relative
populations, mean luminosities, and luminosity dispersions of objects
in the merging sub-clumps.  For example, in the simple two-subclump
model of Zabludoff, Geller, \& Franx (1992), a superposition of two
Gaussian sub-clumps results in $h_3\sim 0$ and $h_4 \sim 0.15$ if the
two populations have identical mean luminosities but the second
sub-clump has a substantially larger luminosity dispersion than the
first clump, while containing a factor of $\sim 5$ times fewer
objects.  Such models are probably too simplistic to realistically
reproduce the formation of the globular cluster halo, although they do
provide some insight into the higher order moments that might be
expected to result from the combination of similarly shaped globular
cluster luminosity distributions in a merger.
\vfill

\section{SUMMARY AND CONCLUSIONS}

Orthogonal Gauss-Hermite functions, which minimize the effects of
outliers in the tails of the observed distribution, show that the GCLF
is well represented by a a Gaussian with $<M_V>=-7.41\pm0.11$ mag and
$\sigma(M_V)=1.24$ mag. Various subsamples of the data can also be well
represented as Gaussians with small higher-order perturbations.  The
dominant higher-order term in the Gauss-Hermite expansion is a
symmetric $h_4 \sim 0.1 $ term.  Subsamples of the data based on
metallicity and Galactocentric radius do not differ from each other
at high levels of statistical significance. In particular, the
luminosity function of metal-poor halo clusters does not appear to
differ markedly from that of metal-rich disk globulars, although there
is some evidence for a weak effect at the $\sim1\sigma$ level.  These
results suggest that, for most applications, a simple Gaussian
description of the GCLF is an adequate representation of the data.
Non-Gaussian representations of the GCLF, such as that resulting from a
t-distribution (Secker 1992) or from the multiple power-law
mass-spectrum model of Harris \& Pudritz (1994), require more free
parameters, and do not result in substantial improvements to the
modelling of the luminosity function. The Harris \& Pudritz (1994)
description of the GCLF is appealing because it results from a
well-defined physical model, but it is it is not yet clear that the
mass-spectrum of giant molecular clouds at the current epoch should be
similar to the mass spectrum of globular cluster progenitors at the
epoch of cluster formation.

\acknowledgements

We thank Roeland van der Marel and Bill Harris for useful comments on
the manuscript, Peter Stetson for interesting discussions on our
analysis methods, and David Duncan for drawing the Figures in this paper.

\clearpage

\begin{figure}
\caption{(Top) The distribution of all 138 Galactic globular clusters in
our sample shown as a histogram, along with the reconstructed
distribution obtained from the Gauss-Hermite decomposition. (Bottom)
The reconstructed distribution (dark solid line), along with the $h_0$
(thin solid line), asymmetric $h_3$ (dashed line), and symmetric $h_4$
(dotted line) components.}
\end{figure}

\begin{figure}
\caption{As for Figure~1, except for the 103 Globular clusters with
$[Fe/H]<-0.8$}
\end{figure}

\begin{figure}
\caption{As for Figure~1, except for the 96 Globular clusters with $3
< R({\rm kpc}) < 30$}
\end{figure}

\begin{figure}
\caption{As for Figure~1, except for the 71 Globular clusters with
$[Fe/H]<-0.8$ \& $3 < R({\rm kpc}) < 30$}
\end{figure}

\begin{figure}
\caption{The 2$\sigma$ confidence regions expected from a Gauss-Hermite
decomposition of a sample of 100 objects.  The upper ellipse shows the
confidence region for a distribution with $h_3=0.05$, $h_4=0.13$ (point
A), corresponding to our low-metallicity subsample of globular
clusters. The lower ellipse shows the confidence region for a perfect
gaussian with $h_3=h_4=0$ (point C).  Point B corresponds to $h_3=0.02,
h_4=0.06$ (our derived parameters for the total sample of 138 globular
clusters).  The confidence regions were estimated from 1000 Monte Carlo
simulations, assuming $M_V=-7.4$ mag and $\sigma(M_V)=1.2$ mag.}
\end{figure}
\vfill
\clearpage

\begin{planotable}{cccccc}
\tablewidth{6.9truein}
\tablecaption{PARAMETERS FOR SUBSAMPLES OF GLOBULAR CLUSTERS}
\tablehead{
   \colhead{Subsample} &
   \colhead{N} &
   \colhead{$<M_V>$} &
   \colhead{$\sigma(M_V)$} &
   \colhead{$h_3$} &
   \colhead{$h_4$}
}

\startdata
All                                      & 138    &  $-7.41\pm 0.11$\ \ \ \ \ \
   & 1.24  &  $0.02\pm0.05$   & $0.06\pm0.05$  \\
$[Fe/H]<-0.8$                            & 103    &  $-7.48\pm 0.11$\ \ \ \ \ \
   & 1.08  &  $0.05\pm0.05$   & $0.13\pm0.05$  \\
$3 < R({\rm kpc}) < 30$                  &  96    &  $-7.46\pm 0.12$\ \ \ \ \ \
   & 1.19  &  $0.02\pm0.06$   & $0.07\pm0.06$  \\
$[Fe/H]<-0.8$ \& $3 < R({\rm kpc}) < 30$ &  71    &  $-7.52\pm 0.12$\ \ \ \ \ \
   & 1.03  &  $-0.01\pm0.07$  & $0.10\pm0.07$  \\
\end{planotable}
\vfill

\end{document}